\documentclass[11pt]{article}
\usepackage{graphicx}
\usepackage{amsmath}
\usepackage{amssymb}
\usepackage{amsxtra}

\usepackage{hyperref}
\usepackage{float}

\title{Mass matrix parametrization for pseudo-Dirac neutrinos}
\author{A. Gorin\\National Research Nuclear University MEPHI\\(Moscow Engineering Physics Institute),\\115409 Moscow, Russia\\INR RAS\\(Institute for Nuclear Research of the Russian Academy of Sciences),\\108840, Troitsk, Russia,\\ e-mail gorin@inr.ru}

\begin{document}
\maketitle

\begin{abstract}
An overview of pseudo-Dirac neutrino framework is given starting from general spinor phenomenology. The framework is then tested by simulation of oscillations for T2K experiment parameters. Two possible derivations \cite{orig} and \cite{orig2} of oscillation parameters are indicated to have the same result.
\end{abstract}


\section{Introduction}

Massive neutrinos directly indicate presence of physics beyond the Standard model (BSM). Precise measurements of neutrino oscillations provide the possibility to probe various BSM theories.

Since the absolute values of neutrino masses are currently beyond direct measurements various experiments are focused on the standard neutrino model ($\nu$SM) oscillation parameters -- square mass differences $\Delta m^2$ and $\delta$-phase.

Some experiments however reported the existence of anomalies in experimental data. These anomalies can find explanation in theories with additional neutrino interactions, most notably the sterile neutrinos.

Recently a number of short-baseline reactor experiments declared an observation of sterile neutrinos with the significance of $3\sigma$. However the observations are not entirely compatible to each other. The matter is under investigation in the ongoing STEREO, PROSPECT, SoLid and Neutrino-4 experiments.
Experimental evidences suggesting sterile neutrino with mass $\sim 1$ eV can be explained in the simplest way in 3+1 neutrino model.

Standard unitary 3+1 data fit suffers from strong tension between MINOS and MINOS+ bound on $\overset{(-)}{\nu_\mu}$ disappearance \cite{nu1} and LSND\&MiniBooNE $\overset{(-)}{\nu_\mu} \rightarrow \overset{(-)}{\nu_e}$ appearance \cite{nu2,nu3}. There are two ways to approach this problem.

First possibility is to consider 3+1 non-unitary mixing scenario \cite{nu4}. It can be used to explain short-baseline disappearance experiments however the anomalies observed in LSND and MiniBooNE experiments \cite{nu5} remain unexplained.

Second possibility is addressing to more than one sterile neutrino. 3+2 scenario can be studied in general framework of 3 active and 3 sterile neutrino.
Here we are probing the pseudo-Dirac scenario with 3 active and 3 sterile neutrinos.

In Section \ref{s:gen} we will describe how pseudo-Dirac neutrinos naturally arise when the neutrino is a composition of Dirac and Majorana spinors. 

In Section \ref{s:mod} we will show that pseudo-Dirac neutrinos can be effectively described by three parameters. Then the mass matrix can be effectively diagonalized which we show using two different approaches. Then we will plot the oscillation probability for pseudo-Dirac scenario against pure Dirac neutrinos for the setup of T2K experiment.

In Section \ref{s:fin} we will discuss what can be further done to address the problem of streile neutrinos and neutrino mass generation.

\section{General spinor formalism}\label{s:gen}

Lagrangian mass term for two spinors $\chi$ and $\eta$ has the form
\begin{equation}
\label{lagr}
\mathcal{L}_{mass} = \frac{1}{2} \begin{pmatrix}
\chi  & \eta 
\end{pmatrix} \mathbf{M} \begin{pmatrix}
\chi  \\ \eta 
\end{pmatrix}
\end{equation}
where mass is given by
$\mathbf{M} =  \begin{pmatrix}
A & M \\ M & B 
\end{pmatrix}$ and $M, A, B$ are 2x2 matrices.

For the most general free field case we can write down ``Weyl-Majorana-Dirac equation''
\begin{equation}
\begin{split}
i\sigma_\mu \partial^\mu \psi_L - \eta_{D,R} m_{D,R} \psi_R - \eta_L m_L (i\sigma_2) \psi^*_L = 0\\
i\bar{\sigma}_\mu \partial^\mu \psi_R - \eta_{D,L} m_{D,L} \psi_L - \eta_R m_R (i\sigma_2) \psi^*_R = 0
\end{split}
\end{equation}
with non-negative mass terms $m$ and phase terms $\eta=e^{i\varphi}$ from unitary group $U(1)$.
Defining  $\tilde{m} = \eta m$ and $\psi_R = \begin{pmatrix}
\psi_1 + i\psi_2  \\ \psi_3 + i\psi_4 
\end{pmatrix}$, $\psi_L = \begin{pmatrix}
\psi_5 + i\psi_6  \\ \psi_7 + i\psi_8 
\end{pmatrix}$ this equation can be transformed into the form \cite{form}:
\begin{equation}
\Box \Phi + \hat{M}^2 \Phi = 0
\end{equation}
where $\Phi =(\psi_1 .. \psi_8)^T$

Now let us illustrate only the simple case $m_{D,L}=m_{D,R}=m_{D}$. For this case general spinor mass matrix is positive semi-definite Hermitian matrix of the form
\begin{equation}
\hat{M}^2 = \begin{pmatrix}
M_R & 0 & 0 & A\\
0& M_R & -A & 0\\
0 & -B & M_L & 0\\
B & 0 & 0 & M_L\\
\end{pmatrix}
\end{equation}

where
$M_R =\begin{pmatrix}
\nu_1 + m_R^2 & -\nu_2\\
\nu_2 & \nu_1 + m_R^2
\end{pmatrix}$, $M_L =\begin{pmatrix}
\nu_1 + m_L^2 & -\nu_2\\
\nu_2 & \nu_1 + m_L^2
\end{pmatrix}$,

$B= \begin{pmatrix}
\mu_1&\mu_2\\
\mu_2&-\mu_1
\end{pmatrix}$, $A= \begin{pmatrix}
k&0\\
0&-k
\end{pmatrix}$
and $\tilde{m}_D \tilde{m}_L + \tilde{m}_D^* \tilde{m}_R = k \geqslant 0$ and moreover $\tilde{m}_D^* \tilde{m}_L + \tilde{m}_D \tilde{m}_R = \mu_1 +i\mu_2$ $\tilde{m}_D^2 = \nu_1 + i\nu_2$
This matrix has four doubly degenerate eigenvalues.
Considering real and positive $m_R$ and $m_D$ and complex $m_L$ we are down to just two eigenvalues.

Now consider $\chi$ and $\eta$ in \ref{lagr} to be the left- and right-handed neutrino fields $\nu_L$ and $\nu_R$. We can work with two Majorana neutrinos if we stipulate $\nu_R=\nu^{'C}_L$. Then
$\mathbf{M} =  \begin{pmatrix}
m_L & m_D \\ m_D & m_R 
\end{pmatrix}$
There are three commonly known special cases for the values of the elements of this matrix.

First case is $m_L=m_R$. In this scenario we have a pair of eigenvalues $m_D \pm m_L$ and mixing angle between $\nu_L$ and $\nu_R$ is given by $tan2\theta = \frac{2m_D}{m_R - m_L} = \frac{\pi}{4}$. No active-sterile oscillations are realized in this case.

Second case is $m_L=m_R=0$. In this scenario we have a pure Dirac neutrino.

Last case is $m_L,m_R \ll m_D$. This scenario is referred to as pseudo-Dirac case.

In general, neutrino can have Majorana and Dirac parts
\begin{equation}
\mathcal{L}^{D+M}_{mass}= \mathcal{L}^D_{mass} + \mathcal{L}^L_{mass} + \mathcal{L}^R_{mass}
\end{equation}
and Dirac neutrino can be represented as two Majorana neutrinos.
Left-handed neutrinos are concerned active while right-handed are sterile i.e. they are singlets under $SU(2)_L \times U(1)_Y$.

For the Pseudo-Dirac neutrino the symmetry of mass matrix is not the symmetry of the weak interaction.
It is easy to obtain Pseudo-Dirac neutrino decomposition
\begin{equation}
\label{dec}
\begin{split}
\psi_{\pm L} = \frac{1}{\sqrt{2}} \begin{pmatrix}
0 \\ \eta_1 \pm i\eta_2 
\end{pmatrix} = \frac{1}{\sqrt{2}} (N_{1L} \pm i N_{2L})\to  \frac{1}{\sqrt{2}} (N_{1L} \pm e^{i\varphi} N_{2L})\\
\psi_{\pm R} = \frac{1}{\sqrt{2}} \begin{pmatrix}
-i\sigma^2 (\eta^*_1 \pm i\eta^*_2)  \\ 0
\end{pmatrix} = \frac{1}{\sqrt{2}} (N_{1L}^C \pm i N_{2L}^C) \to \frac{1}{\sqrt{2}} (N_{1L}^C \pm e^{i\varphi} N_{2L}^C)
\end{split}
\end{equation}
for a pair of almost degenerate mass Majorana neutrino with opposite CP sign and lepton number not being conserved in higher order weak interaction.

Because of the small value of mass matrix distortions the mixing angle between two Majorana neutrinos is $\sim \frac{\pi}{4}$.

\section{Modeling}\label{s:mod}
\subsection{Mass matrix diagonalization}

For chirality preserving processes it is suffice to diagonalize $M^{\dagger} M$. We will now consider two possibilities -- $M^2$ and $M$ diagonalization and show that in the leading order they provide the same result for pseudo-Dirac neutrinos.

In general, 6x6 mass matrix diagonalization gives 15 mixing angles, multiple violating CP phases and 6 eigenvalues.
Under Pseudo-Dirac assumption this can be approximated by ordinary 3x3 Pontecorvo-Maki-Nakagawa-Sakata (PMNS) matrix \cite{orig}.
\begin{equation}
M^{\dagger} M \simeq \begin{pmatrix}
m^{\dagger}_D m_D & m^*_L m^T_D + m^{\dagger}_D m^*_R \\ m^*_D m_L + m_R m_D & m^*_D m^T_D
\end{pmatrix}
\end{equation}
consider bi-unitary transformation $U^{\dagger}_R m_D U_L = diag(m_1,m_2,m_3)=m$ then $V=\begin{pmatrix}
U_L & 0\\
0 & U^*_R
\end{pmatrix}$ and
\begin{equation}
V^{\dagger} (M^{\dagger} M) V =\begin{pmatrix}
m^2 & U^{\dagger}_L m^{\dagger}_L U^*_L m + m U^{\dagger}_R m^*_R U^*_R\\
m U^T_L m_L U_L +U^T_R m_R U_R m & m^2
\end{pmatrix}
\end{equation}

If we completely ignore off-diagonal parts then it is just Dirac scenario with doubly-degenerate eigenvalues.
Otherwise in the first order approximation each pair takes the form
$\begin{pmatrix}
m^2_i & \epsilon^*_i m_i\\
 \epsilon_i m_i & m^2_i 
\end{pmatrix}$

Now we obtain 6 mass eigenstates 
$\nu_{iS}=\frac{1}{\sqrt{2}}(\nu_{iL}+e^{i\varphi_i}\nu_{iR})$ $\nu_{iA}=\frac{1}{i\sqrt{2}}(\nu_{iL}-e^{i\varphi_i}\nu_{iR})$
such that $e^{i\varphi_i} = \frac{\epsilon_i}{|\epsilon_i|}$ for decomposition \ref{dec} and
mass eigenvalues given by $m^2_{iS,A}=m_i^2 \pm \epsilon_i m_i$.

Another method for diagonalization $M$ itself is completely removing left-handed Majorana spinor part of the Dirac one --
mass matrix takes the form $M=\begin{pmatrix}
\mathbf{0} & m^{'}_D\\
m^{'}_D & M_s
\end{pmatrix}$
In \cite{orig2} it is shown that the appropriate diagonalizing transformation is given in form
\begin{equation}
V= \frac{1}{\sqrt{2}} \begin{pmatrix}
U^{\dagger} &1\\
U & 1
\end{pmatrix} \begin{pmatrix}
1 & \delta \\
-\delta^{\dagger} & 1
\end{pmatrix}
\end{equation}
where U diagonalizes $m^{'}_D$ and $\delta = U(\epsilon /2 + \varepsilon)$,  $\varepsilon^T = -\varepsilon$ and $M_s = 2\epsilon m_D - \varepsilon m_D + m_D \varepsilon$.
This produces\\
$M=V^{\dagger}mV$ where $m=\begin{pmatrix}
m_D (1+\epsilon) & 0\\
0 & -m_D (1-\epsilon)
\end{pmatrix}$

Now $m^2$ in the leading order have the eigenvalues  $m^2_i \pm \epsilon^{'}_i m_i$ which are the same as in the previous case.

\subsection{Probing the pseudo-Dirac scenario}

With these eigenvalues we can write down the oscillation probability in terms of ordinary PMNS matrix. Assume that mass eigenvalues splitting for pseudo-Dirac neutrino is given by $m^2_{iS,A}=m_i^2 \pm \epsilon_i m_i$. Using the results from \cite{orig} it is easy to model $\nu_\mu \rightarrow \nu_e$ oscillation probability which is

\begin{equation}
P(\nu_\alpha \rightarrow \nu_\beta)=\frac{1}{4}\left | \sum_{j=1}^{3} U_{\beta j} (e^{i \frac{m^2_{jS}}{2E}t}+e^{i \frac{m^2_{jA}}{2E}t})U^*_{\alpha j} \right |^2
\end{equation}

To illustrate potentially observable differences between Dirac and pseudo-Dirac scenario we will simulate oscillations for T2K experiment parameters:
  \begin{itemize}
  \item
    $L=295$ km and $E \leq 2$ GeV.
  \item
    $\delta = - \frac{\pi}{2}$ and $sin^2 \theta_{12} = 0.307 \; sin^2 \theta_{23} = 0.5 \; sin^2 \theta_{13}=0.218$.
  \item
    $\Delta m^2_{12}=7.53 \cdot 10^{-5} \text{eV}^2 \; \Delta m^2_{23}=2.44 \cdot 10^{-3} \text{eV}^2$.
    \item
    normal mass hierarchy.
  \end{itemize}

\begin{figure}[H]
\centering
\includegraphics[scale=0.6]{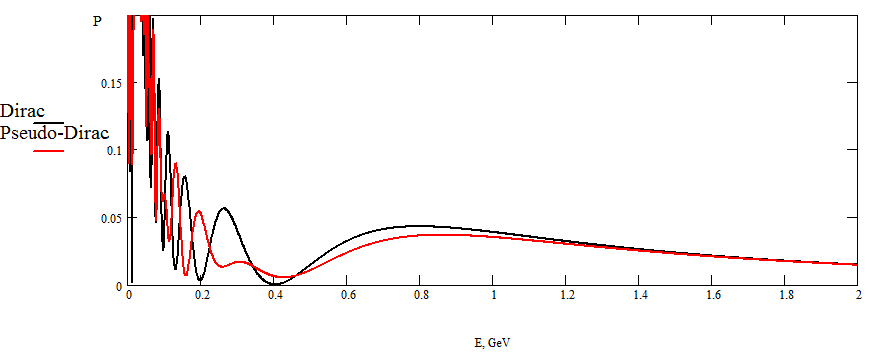}
\caption{Pseudo-Dirac neutrino $\nu_\mu \rightarrow \nu_e$ oscillation probability compared to pure Dirac scenario for T2K experiment parameters and naive assumptions for pseudo-Dirac mass eigenvalues.}
\label{fig}
\end{figure}

This allows us to probe the impact of small Majorana additives.  Please also note that energy spectrum now depends on the absolute mass of neutrino because of the splitting.  First we will model the situation where $\epsilon_i = 0.1$, Fig. \ref{fig}.

 Please note that neutrino beam in T2K experiment has energy distribution with maximum at $0.6$ GeV and almost all neutrinos have energy in the interval $0.5 \div 1$ GeV. So we cannot make any assumptions considering pseudo-Dirac neutrinos using only T2K data.

\begin{figure}[H]
\centering
\includegraphics[scale=0.6]{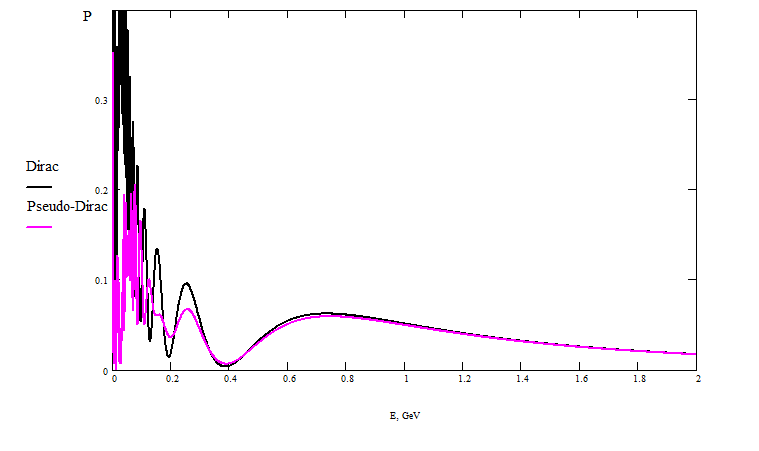} 
\caption{Pseudo-Dirac neutrino $\nu_\mu \rightarrow \nu_e$ oscillation probability compared to pure Dirac scenario for T2K experiment parameters for more realistic mass splittings.}
\label{fig2}
\end{figure}

Let us illustrate the difference in energy spectrum for more realistic $\epsilon_i$ parameters.
In Fig. \ref{fig2} we have taken $m_1=0.01$ eV, $\epsilon_1=2.6 \cdot 10^{-3}$,$\epsilon_2=4.0 \cdot 10^{-3}$ and $\epsilon_3=5.0 \cdot 10^{-3}$ proportional to mass squares differences.

\section{Discussion and Conclusion}\label{s:fin}

Now we are in the situation where combined experimental data from atmospheric, reactor and accelerator neutrino experiments is in good agreement with 3 active neutrino model for the first three oscillation peaks. Upcoming experiments can provide more experimental data thus clarifying the situation.

Long-baseline experiments can provide precise values of $\nu$SM oscillation parameters and provide enough data to determine the neutrino mass hierarchy.

Short-baseline experiments can either improve their statistics and cancel out all anomalies or successfully approve that the $\nu$SM needs expansion.

Using precise $\beta$-decay and K-capture measurements it would be arguably possible to measure neutrino masses directly or at least put a constraints on them.

$\beta \beta$ and $0\nu \beta \beta$ observations as well as atmospheric, solar, galactic and extra-galactic neutrino experiments are important for probing different neutrino mass generation mechanisms.
  
It is also important to consider theoretical models for processes in early Universe -- the constraints from these models are generally less strict than from direct observations but still helpful either for a cross-checking or for limiting the potential of exotic mass generation and mixing models.

Here we presented the derivation of pseudo-Dirac neutrino from general spinor formalism.

For the parameters of T2K experiment the probability of $\nu_\mu \rightarrow \nu_e$ oscillation was modeled. The current setup of the experiment however is not sensitive to differences in Dirac and pseudo-Dirac oscillations.

It was shown that in the leading order approximation PD neutrino can be effectively described by three $\epsilon$ parameters of mass splitting -- it is valid for $M^2$ and $M$ diagonalization.

There are questions arising naturally in the context of neutrino mass generation mechanism. 

First question is whether it is suffice to consider pseudo-Dirac neutrino to fit observations or general framework is needed? This question will be addressed by the future observations.

Second question is about the compatibility of particular mass generation mechanism with pseudo-Dirac scenario in particular and it's rigidity to possible observational data as a whole. Which mechanisms are the best candidates, Yukawa coupling or multiple scalar fields (like in Zee model) or maybe even geometric models of mass generation?


\section*{Acknowledgements}
I am grateful to M.Yu. Khlopov for an invitation to XXII Bled Workshop and to the organizing committee of the Workshop for an opportunity to make a talk via internet.



\begin{thebibliography}{99}
\bibitem{form}  A. Aste: Weyl, Majorana and Dirac fields from a unified perspective, 
 	Symmetry, \textbf{8(9)}, 87 (2016).
 	
\bibitem{nu1} P. Adamson et al: Search for sterile neutrinos in MINOS and MINOS+ using a two-detector fit,
	 Phys. Rev. Lett. \textbf{122}, 091803 (2019).
	 
 \bibitem{nu2} S. Gariazzo, C. Giunti, M. Laveder, Y.F. Li: Model-Independent $\bar{\nu}_e$ Short-Baseline Oscillations from Reactor Spectral Ratios,
	 Phys. Lett. B \textbf{782}, 13 (2018).
	 
 \bibitem{nu3} M. Dentler et al: Updated global analysis of neutrino oscillations in the presence of eV-scale sterile neutrinos,
	 J. High Energ. Phys. \textbf{2018}: 10  (2018).
	 
\bibitem{nu4} C. Giunti: Short-baseline neutrino oscillations with 3+1 non-unitary mixing,
	 Phys. Lett. B \textbf{795}, 236  (2019).
	 
\bibitem{nu5} MiniBooNE Collaboration, A. A. Aguilar-Arevalo et al: Significant Excess of ElectronLike Events in the MiniBooNE Short-Baseline Neutrino Experiment,
	 Phys. Rev. Lett. \textbf{121}, 221801 (2018).
	 
\bibitem{orig} M. Kobayashi, C.S. Lim: Pseudo-Dirac Scenario for Neutrino Oscillations,
	  	Phys. Rev. D \textbf{64}, 013003 (2001).

\bibitem{orig2} A. de Gouvea, W.C. Huang, J. Jenkins: Pseudo-Dirac Neutrinos in the New Standard Model,
	 Phys. Rev. D \textbf{80}, 073007 (2009).



\end{thebibliography}
\end{document}